# GTA - An ATSP Method:
# Shifting the Bottleneck from Algorithm to RAM

Wissam Nakhle [1,2]

*[1]Department of Mechanical, Industrial and Aerospace Eng. (MIAE), Concordia University, Montreal, Canada*

*[2]Department of Mechatronics Eng., American University of Science and Technology Beirut, Lebanon*

**ABSTRACT**

We present a scalable, high-performance algorithm that deterministically solves large-scale instances of the Traveling Salesman problem (in its asymmetric version, ATSP) to optimality using commercially available computing hardware. By combining an efficient heuristic warm start, capable of achieving near-optimality within seconds in some cases, with a subtour elimination strategy that removes the need for traditional MTZ constraints, our approach consistently resolves instances up to 5,000 nodes (approximately 25 million binary variables) in record time on widely accessible computers, with eight logical processors. We demonstrate reproducible results with convergence rates comparable to those of high-performance computing frameworks. Real-time iteration tracking and an adaptable interface allow seamless integration into scheduling workflows in logistics, bioinformatics, and astronomy. Designed to streamline solutions to large-scale TSP problems across disciplines, our approach is benchmarked against widely used public datasets, offering a deterministic, resource-efficient alternative to conventional solvers that rely on supercomputing hardware. Our GTA (Gurobi Tabu Algorithm) algorithm is a fundamental shift of TSP solution bottleneck from algorithmic complexity to the underlying hardware (RAM and system memory), which is a highly desirable characteristic.

## INTRODUCTION

Recent advancements in combinatorial optimization offer new perspectives on the Traveling Salesman Problem [1]. With diverse applications in logistics and routing, including extensions to the location-routing problem [2], and the traffic assignment problem [3], the Traveling Salesman Problem (TSP) is still one of the most frequently studied optimization problems [4]. In astronomy, TSP finds applications in areas such as telescope observation scheduling [5], using mixed-integer linear programming [6]. Integer programming has also proven crucial for relay scheduling of telescope arrays [7], and for large complex astronomical surveys, such as the Rubin Observatory [8]. Robust algorithms for quality assessment of genome assemblies are essential [12]. In genomics, TSP is deeply embedded in applications involving DNA sequencing [9]. Pre-processing reduction methods may help manage this complexity [10], but algorithms for DNA sequences and assemblies inevitably rely on a TSP approach in addressing the most complex genomic applications [11].

The optimal solution to TSP determines the unique and complete sequence that visits each node exactly once while minimizing total cost, time, or distance. Although small instances of TSP (up to a threshold) can be solved relatively quickly, solution complexity grows rapidly as problem size increases. Addressing the computational complexity of TSP requires a comprehensive understanding of the optimization parameters involved [13]. More recent approaches include a polynomial time approximation, providing valuable insight into the theoretical framework and the complexity of TSP solutions [14]. Heuristics such as the 2-Opt (amongst other heuristics) can approximate TSP solutions and produce sub-optimal results several orders of magnitude faster than deterministic approaches [15]. TSP is an NP-hard (Non-Polynomial) problem and becomes increasingly more challenging to solve as the number of nodes grows, especially when deterministic solutions with 0% optimality gaps are required. Although extensive literature on the Traveling Salesman Problem and its variations provides a broad context and a benchmark for research, and while significant efforts in heuristic, exact, and learning-based methods have been explored, no existing framework ensures optimality, repeatability, and scalability for large-scale TSP instances without requiring supercomputing and/or computer clusters.

Solving TSP problems with variants such as time window requirements have been explored using various algorithms [17]. Q-learning is a metaheuristic method for solving TSPs, including delivery system problems [18], but it remains limited as urban delivery problems require an exact and deterministic approach due to conflicting time window requirements [19]. Other methods such as Kullback–Leibler minimization has also been explored [20]. The minimum spanning tree problem, another TSP-related combinatorial optimization problem, has also been extensively reviewed for the similarity of its formulations and solution procedure to TSPs [21]. Branch and bound methods and related algorithms (such as interior-point barrier methods) remain the most common techniques for solving traveling salesman problems [22], even though significant resources have been recently allocated to the use of machine learning in solving combinatorial optimization problems [23].

Existing literature attempts to address the NP-hard nature of TSP through various approaches: A) MIP exact solvers such as Concorde, CPLEX and Gurobi, that predominantly use a branch-and-cut approach (combined with other related algorithms) to achieve 0% optimality gaps, at the cost of higher time complexity and memory use as problem size increases. B) Heuristic algorithms, including 2-opt/3-opt, simulated annealing [24], tabu search, genetic algorithm and nature-inspired algorithms [25], hyper-heuristics [26], iterated local search [27], and average-case subquadratic procedures [28], all of which aim to improve initial feasible solutions and iterate quickly, but sacrifice exactness and reproducibility, while providing modest optimality gaps, even for NP-complete TSP problems [29]. C) With the rise of artificial intelligence, neural combinatorial optimisation and reinforcement learning (RL) for TSP have been widely explored. Convolutional neural networks (CNN) [30, 31], reinforcement learning (RL) [32, 33, 34] and machine learning (ML) [35, 36] have recently gained traction, and while promising in various fields, they still struggle to consistently achieve 0% optimality gaps and often lack repeatability, due to their dependence on large training datasets [37] and sensitive parameter tuning [38, 39, 40]. New benchmarks like the GalaxyTSP, with its billion-node benchmark for TSP, pushes the boundaries for testing TSP algorithms and provides valuable insights [41].

Surrogate-assisted [42, 43] computational algorithms as well as parallel computing and energy-efficient Ising machines have also been explored to address computational challenges in solving large-scale TSP instances [44, 45], particularly for expensive constrained TSP problems. Deep learning methods have also been designed to handle more complex TSP cases, involving refueling constraints [46, 47]. Improving the performance of heuristics such as the Ant Colony heuristic [48] and other quantum-based models [49, 50] are continually being explored, and their performance regularly evaluated. A more recent and promising, yet still not fully explored, is the use of quantum computing in solving large TSPs [49, 50]. Research efforts have already started to show how quantum models could accelerate combinatorial optimisation and offer new unprecedented computing capabilities for solving complex problems [49, 50]. Quantum computing approaches to TSP are still in an early stage and, while promising, they are currently not able to solve real-world TSP instances with the consistency and precision required. All these contemporary methods offer new perspectives on addressing the computational complexity inherent to NP-Hard TSP problems, but they all struggle to achieve true optimality with 0% gaps (within a practical timeframe) and fail to consistently handle large-scale, real-world TSP instances. There is a constant need to bridge this gap in performance and solutions to TSP-related problems. Most methods either sacrifice speed for optimality or produce fast results at the expense of exactness and reproducibility.

Previous research has explored a wide spectrum of methods, all of which remain widely dominated by heuristic and deterministic methods, with limited success in balancing solution time, quality and scalability. Consistently generating deterministic solutions for large-scale TSP instances, without the need for supercomputing or computer clusters, and within accelerated timeframes, remains uncharted territory, a largely unexplored frontier and a critical gap in research. This study presents a novel optimisation approach to TSP that directly addresses these limitations. To bridge this gap in research, reconciling optimality and time complexity, our algorithm delivers a robust optimization framework capable of deterministically solving large-scale TSP instances in unprecedented runtimes using standard desktop machines. Addressing this complex computational problem is vital for scientific and industrial progress.

This paper introduces a TSP solver that is both deterministic and scalable, producing optimal results within remarkable runtimes on widely accessible hardware (4 to 8 cores and 8 to 12 logical processors). Our high-performance solver is first and foremost designed to be exact, scalable, and capable of solving large-scale TSP instances on standard hardware in remarkably short timeframes. The primary goal is to achieve both precision and speed in deterministically solving complex and large TSPs. This paper presents a high-performance, exact and ultra-fast streamlined solver that can be simply deployed on standard commercial hardware (i.e. PCs and desktop machines with moderate capabilities), combining intelligent heuristic warm starts, that guarantees less than 5% gaps in under 10 seconds (for the largest problems) with an MTZ-free subtour elimination algorithm, all integrated into a customised MIP configuration using Gurobi as the main solver. Our algorithm guarantees exactness and repeatability, making it suitable to solve other NP-hard combinatorial problems, supported by a streamlined user interface and real-time graphical progress across iterations, with no compromise to performance. Our method combines the robustness of commercially available mixed-integer programming solvers (MIP – specifically Gurobi), with strategic heuristic warm starts, yielding initial solutions with gaps ranging from 1.5% to 5% within seconds. Eliminating MTZ constraints, and integrating an MTZ-free formulation with a fast subtour elimination function, all allows our algorithm to achieve unprecedented computational efficiency and reliable performance on standard processors. The solver's unique performance is enabled by these factors: a customised MIP strategy, a high-quality heuristic warm start, and an MTZ-free subtour elimination function. These key elements all contribute significantly to the solver's overall efficiency.

On commercial PCs equipped with 8-core processors, our algorithm generates solutions with a 0% optimality gap across instances of over 5,000 nodes in under 30 minutes (4000 nodes are solved in less than 600sec, or ~10min). Such practical results have broad implications for real-world applications and can be tailored to meet interdisciplinary requirements across a variety of fields. Interdisciplinary TSP cases are discussed, including logistics, routing, genome sequencing/assembly, and telescope observation planning. This innovative integration of various elements into our method, ensures reliable performance across diverse problem scales. With its accurate, predictable, and scalable decision-making framework, this paper advances TSP-related optimisation problems and offers both a versatile and practical algorithm applicable across interdisciplinary scientific and industrial fields.

A milestone in TSP research resides at the frontier of applying exact methods to large instances, which are otherwise so far solved using supercomputers. Our method directly challenges the need for supercomputing with Structural Enhancements of the standard TSP model via MTZ-free subtour elimination using a Hybrid approach with an extremely efficient Tabu Search warm start coupled with a Gurobi MIP solver. Our current implementation extends beyond the standard TSP problem and offers variants that include non-standard TSP constraints that are critical in various interdisciplinary fields. Our algorithm solves TSPs with Asymmetric Costs (ATSP) by default, where the cost of going from node A to node B can be different from that of going from node B to node A. A brief performance comparison between symmetric and asymmetric solutions is provided. Because of this ability to solve ATSPs in unprecedented runtimes, our work is relevant to genomic applications with direction-dependent solution path, such as in DNA sequencing. Time windows, visibility, and position constraints are also available using our algorithm. One of its variants extends to include time window constraints using earliest and latest arrival, as well as service time parameters. This variant is of particular interest to telescope scheduling and observations of celestial targets, as well as a variety of logistics-related problems such as vehicle routing. Visibility constraints may be modeled with excessively high costs when the directions of interest are not observable. Further, if celestial observations have different scientific values, these could be modeled using assignment of priority weights, i.e., weight coefficients integrated directly into the objective function.

Our current model is readily designed to handle asymmetric cases (ATSP), and its variants extend to include time windows constraint. While the algorithm only handles time-independent costs, the model can easily be modified to include time-dependent coefficients. However, we present a thorough analysis of the impact of changing cost parameters on the algorithm's performance and thus its reproducibility and reliability (by applying different seed values to randomly generate the cost matrix). Our efficient and integrative framework is appropriate for exact optimisation in a variety of fields that typically rely on approximations or heuristics, because of our algorithm's ability to maintain a record solution time with 0% optimality gaps for increasing problem size and complexity. We present the results of our effective technique for resolving TSP problems, with potential applications in a variety of fields, such as logistics, genomics, and astronomy.

## RESULTS

Our objective is to overcome traditional trade-offs between accuracy, speed, reproducibility, and adaptability to various TSP-related problems, particularly for large-scale instances. Using a Python-based modular architecture allows us to adapt to TSP variants, and enables users to easily experiment with our methods. In the context of high-performance combinatorial optimisation, and for every run, a dynamic cost matrix is randomly generated. The cost matrix is asymmetric and enables solving asymmetric TSPs (ATSP), which is critical for real-world applications. As a first step, cost values are randomly scaled between 1 and 10 as integers, and diagonal entries are assigned a high value. A seed value is chosen to generate the random cost matrix, and our method's reproducibility is verified in part using different random matrices generated by varying the seed value.

Gurobi is used as the main MIP solver, subtour elimination is guaranteed by using lazy constraint callbacks and the overall performance of our method includes a comparison with the use of traditional Miller-Tucker-Zemlin (MTZ) constraints. The full route, total cost, number of iterations, nodes explored, and runtimes are all recorded to assess performance across all instances, including instances exceeding 5,000 nodes. The main heuristic providing a warm start is a variant of Tabu Search, which was designed from the ground up. It is intended to produce high-quality initial feasible solutions quickly and particularly for complex and large instances. An important element of our customized warm start heuristic includes a quick and near-optimal initial tour produced within seconds and under 5% gaps via a greedy nearest-neighbor heuristic. The user may choose to reverse part of the tour (based on how well Gurobi proceeds from the warm start) using a 3-opt operation to provide further improvements to the warm start solution. This allows for more extensive neighborhood exploration compared to 2-opt. One significant novelty in this work is our highly customized hybrid approach along with the use of lazy constraints. We get a high-quality initial tour within seconds (gaps typically less than 5%), which is then converted to a robust warm start for Gurobi, which proceeds by solving an MTZ-free MIP. Gurobi's search space is greatly reduced by this warm start, allowing it to focus on pruning, and searching in the direction that swiftly produces optimality. This approach has demonstrated exceptional performance on large-scale instances (e.g., 5,000+ cities), achieving 0% optimality gaps in unparalleled time on standard machines, where Tabu Search alone could not achieve 0% gaps or close in on the optimal route, and Gurobi alone would take extremely long runtimes even if MTZ-free subtour eliminations are used, reaching over 24hrs in some cases.

The streamlined user interface allows for real-time iteration tracking, and route density visualization. The number of nodes, i.e., problem magnitude, a seed value, the algorithm choice can all be set by the user, the interface then displays optimal path/route, total cost, runtime in seconds, solver-specific metadata, such as solver configuration, nodes investigated, iterations, and various other parameters can be retrieved. This interactive design facilitates experimentation, as well as prototyping and research of various TSP-related problems.

## COMPLEXITY & PERFORMANCE

We first compare performance to currently available approaches (shown in Table 1), including NP-Hard combinatorial complexities, such as Brute-Force and Dynamic Programming search for exact algorithms, the Christofides, N-Opt, and greedy heuristics, available TSP data (Concorde and TSPLIB), as well as our own approach, using the Gurobi solver with lazy constraints.

**Table 1: An Overview Performance Comparison Against our Gurobi/Tabu GTA Algorithm**

| **Algorithm** | **Complexity - O** | **Optimality Gap %** | **($10^{-5}$) x 5000 Nodes** |
|---|---|---|---|
| *GTA** (Gurobi/Tabu)* | $N^{2.01}$ - $N^{2.03}$ | *0* | *350 – 850 sec* |
| *Gurobi* (Lazy Constraints)* | $N^{2.1}$ - $N^{2.2}$ | *0* | *3750 – 6750 sec* |
| Concorde [51] | rl1304 / vm1084 | 0% (**1304/1084 nodes**) | **103.01 sec / 234.66 sec** |
| TSPLIB [52, 53] | fnl 4461 | 0% (4461nodes) | 182566 sec |
| Brute Force [54, 55] | N! x N | 0 | Very Large |
| Dynamic (Held-Karp) [56] | $N^2 \times 2^N$ | 0 | Very Large |
| N-Opt [57] | $N^2 \times \log(N)$ | 50% or more | 924.74 sec (or more) |
| Greedy [58] | $N^2 \times \log(N)$ | Variable | 924.74 sec (or more) |
| Christofides [59] | $N^3$ | 50% Guarantee | ~1 250 000 sec |

The Traveling Salesman Problem (TSP) has long been a computational challenge due to its NP-hardness. However, by overcoming performance standards and offering provable and consistent optimality in quadratic polynomial time, our hybrid Gurobi/Tabu (GTA) algorithm breaks the barrier between the need for deterministic optimality (down to 0% gaps) and the need for quick practical processing times. GTA is a leap forward in solving TSPs and related problems, because of its intelligent warm start and its optimality-focused MIP solver. It would take unrealistic time for Gurobi and any other conventional and popular solver (Concorde and Mosek) to start their search from a zero-point or a naïve feasible solution, considered as highly distant for a starting point. As shown in Figure 1, GTA becomes a more precise solver, similar to other MIP solvers, due to the critical narrowing of the search space provided by a single step Tabu Search initial traversal as a 'warm start' for Gurobi.

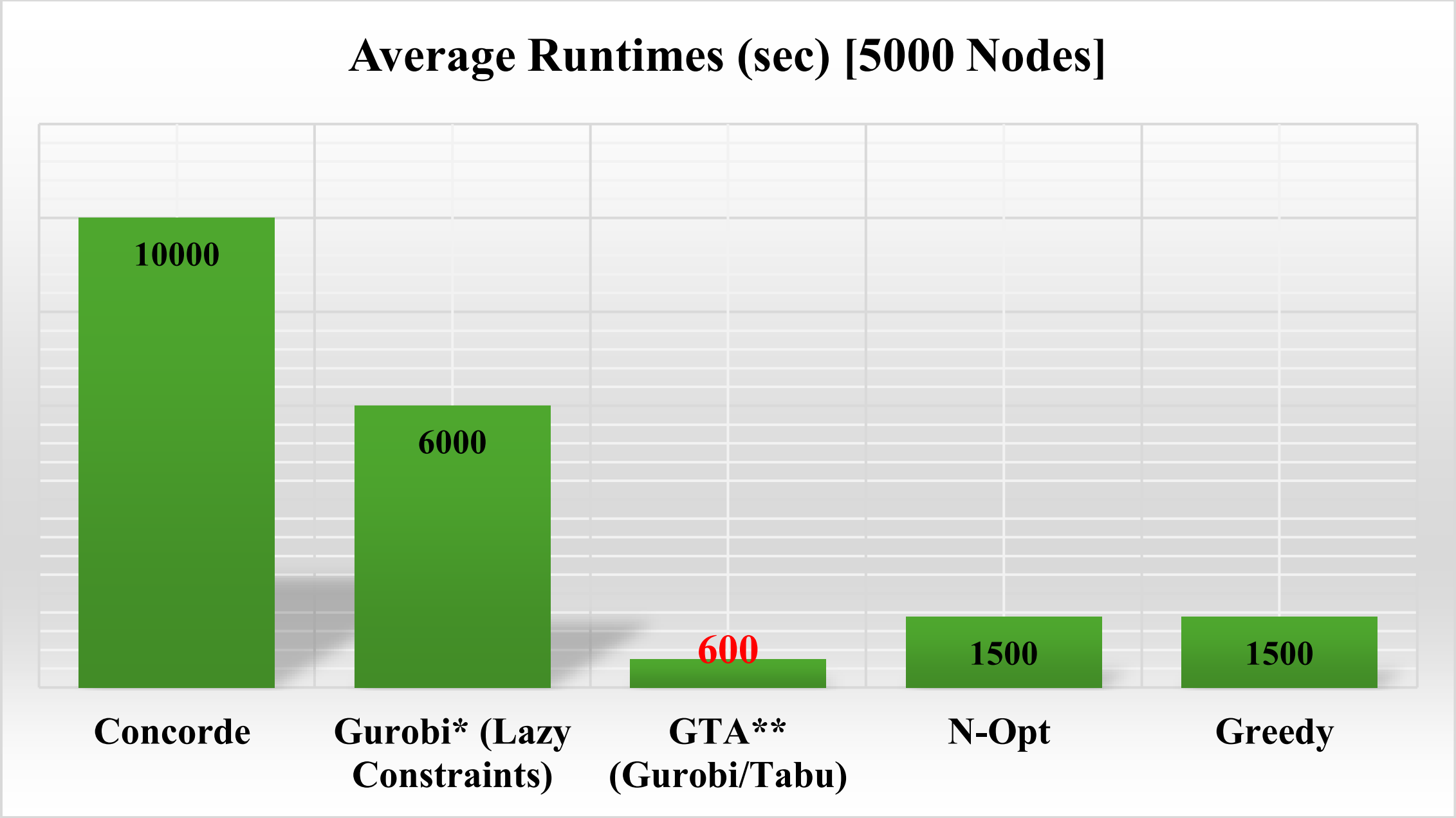

**Figure 1: Comparison of Average Runtimes**

GTA's approach is significantly different from traditional deterministic TSP solvers, which have been criticized for their trade-offs. GTA is a novel technique for TSP optimization that achieves 0% optimality gaps and maintains its consistency for various cost matrices and in large-scale instances of 5,000 nodes and more. This is a significant improvement over standard TSP algorithms which can either only approximate the optimal path in a few minutes or require near-astronomical times to reach 0% gaps, as shown in Figure 2. The GTA algorithm, with its near $N^2$ complexity, advances interdisciplinary TSP research by offering the speed of heuristics without compromising the accuracy of exact algorithms.

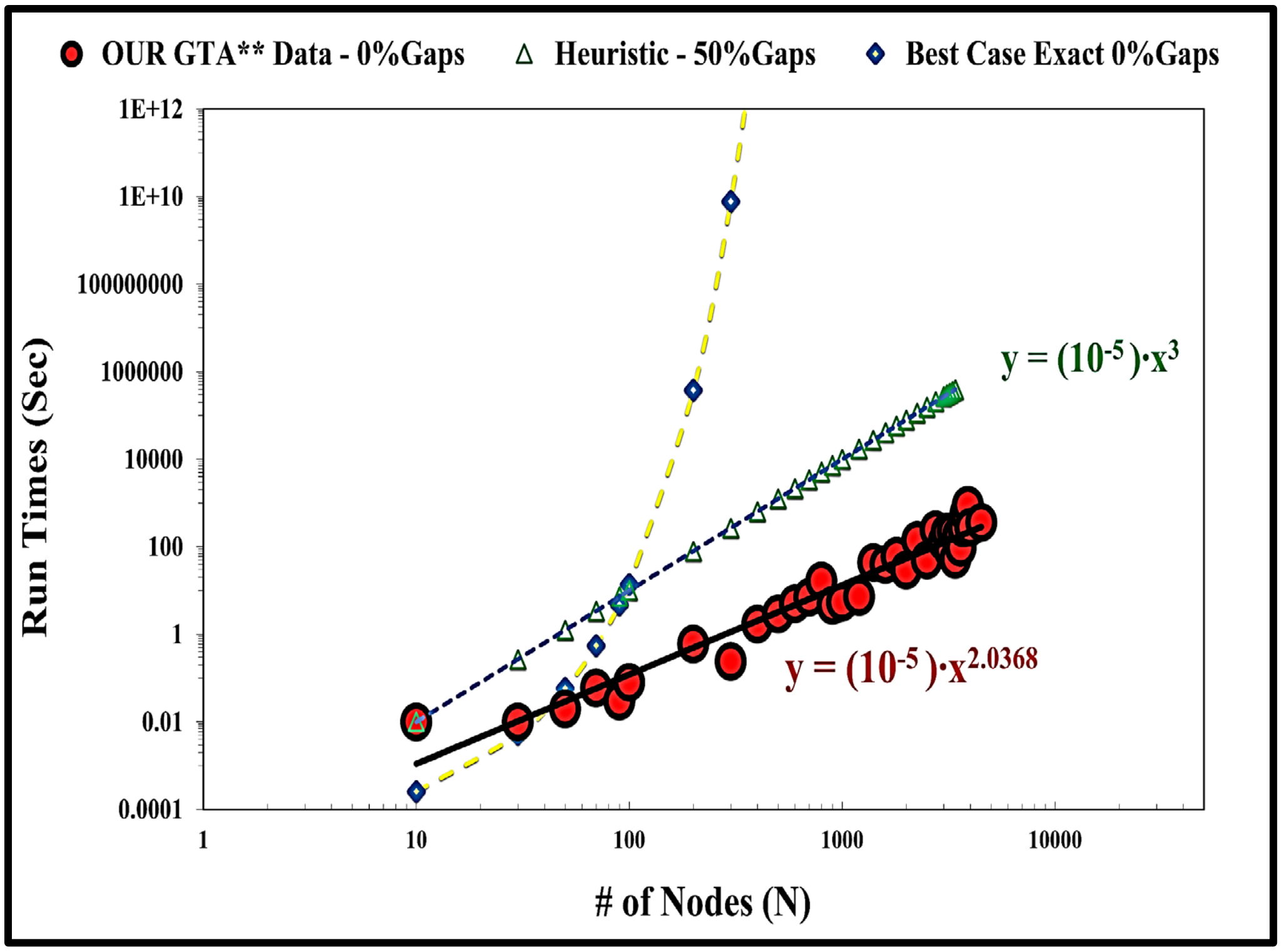

a) Log-Log Scale

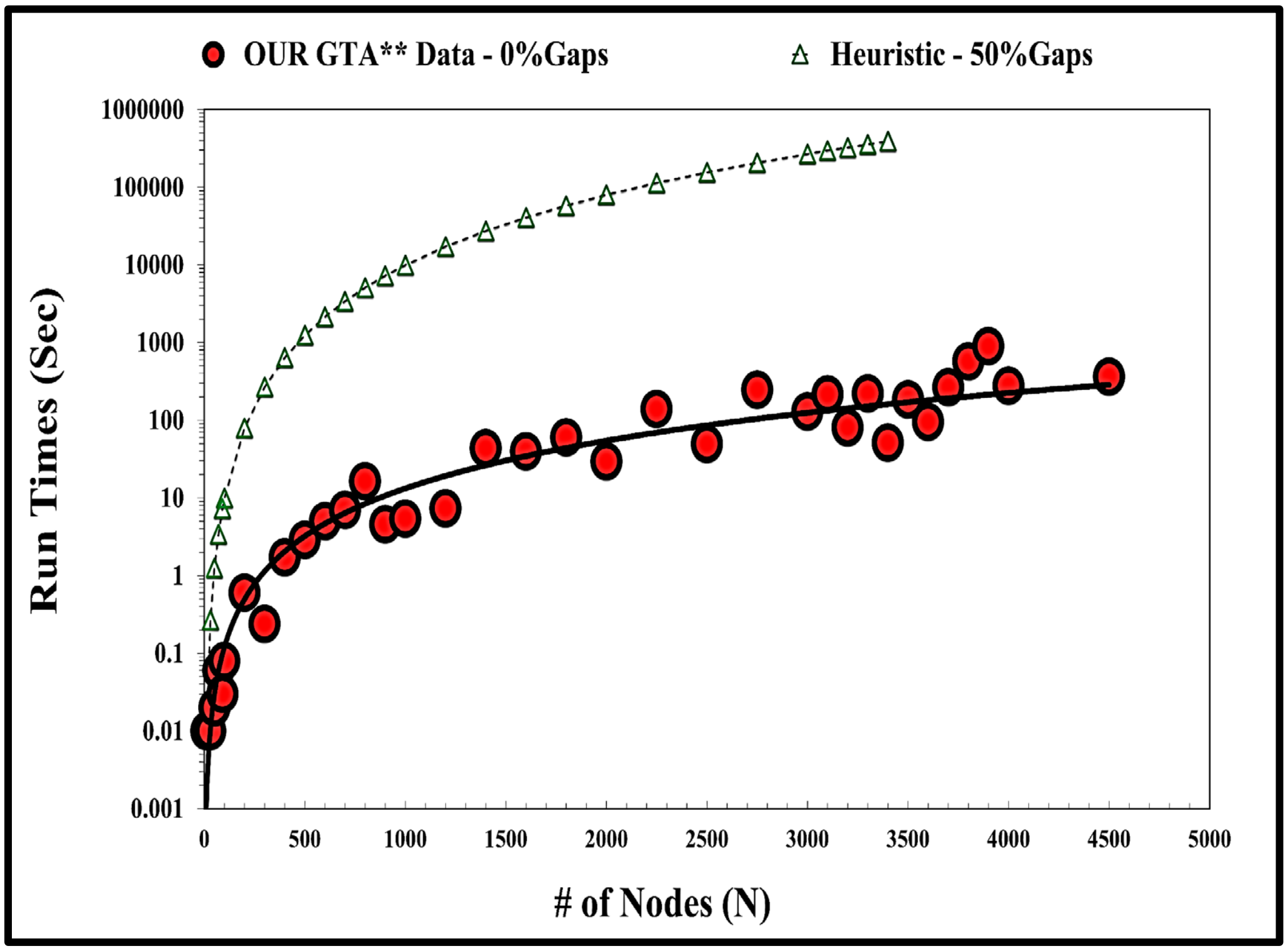

b) Lin-Log Scale

**Figure 2: Performance and Complexity Comparison Against Data From GTA**

Our GTA algorithm's complexity and performance are contrasted with both conventional deterministic and heuristic approaches in Figure 2. The linear plot in Figure 2b indicates a sharp rise in runtime even for a heuristic with 50% gaps, and the log-log plot in Figure 2a shows a clear overshoot in runtimes of exact solvers as the number of nodes increases. Figure 2 demonstrates our GTA's scalability and superior performance over conventional methods, emphasizing its efficiency as problem size grows. Brute Force and Dynamic Programming algorithms are computationally expensive with long and unrealistic runtimes, while heuristics like N-Opt and Greedy heuristics offer faster execution but significantly compromise optimality. Existing exact solvers like Concorde or Mosek, and their related published data (TSPLIB) cannot handle large instances within reasonable time. Even when our GTA is compared with our own Gurobi solver (in Figure 3) with no warm start but with lazy constraints instead of MTZ (and with MTZ), results confirm that without GTA, higher complexity and longer runtimes occur.

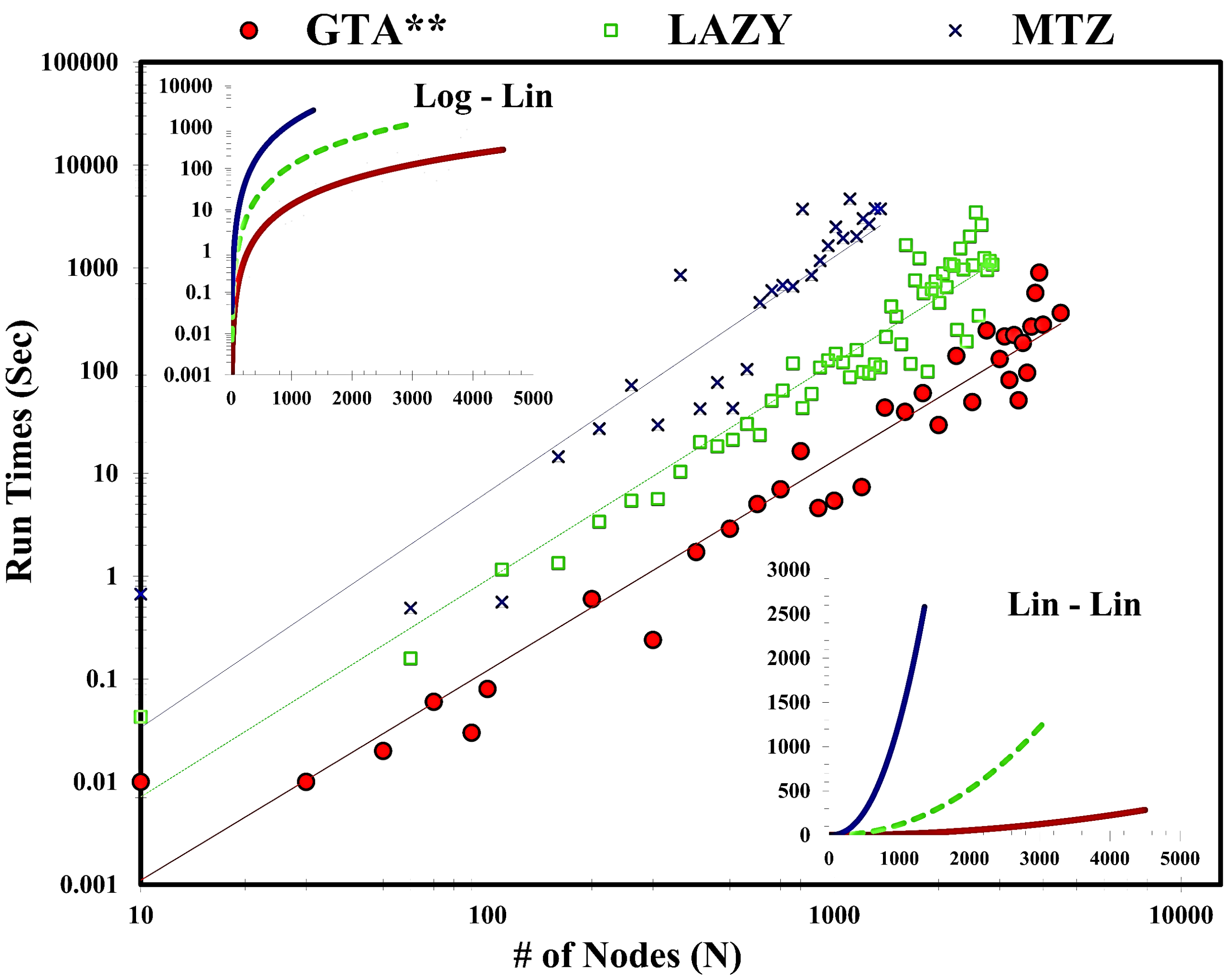

**Figure 3: GTA Performance and Complexity Comparison Against Our Own Lazy and Brute-Force MTZ Solving**

## NETWORK COMPLEXITY VISUALIZATION

TSP solutions range from simple routes to complex networks, as shown in Figure 3, and demonstrated by three images for 10, 100, and 1000 nodes. The first route for N = 10 displays a simple path with ten nodes, the second for N = 100 shows a more complex network with tighter connections, while the third for N = 1000 nodes and the fourth for N = 2000 nodes, reveals the complexity of the TSP network. When the optimal path becomes a complex, closely spaced network of red and black lines, a dense graph is created. Figure 3 illustrates how TSP complexity increases with N, which can be seamlessly vizualised. GTA solves thousands of nodes deterministically with 0% optimality gaps within minutes, demonstrating exceptional performance with an $O(N^2)$ complexity. GTA uses intelligent pathfinding and pruning, to swiftly converge to the optimal path. The warm start first provides a high-quality route, allowing Gurobi to then prune and eliminate many suboptimal branches from the search space. The algorithm's ability to identify the optimal sequence that minimizes costs and that generates a single and elegant optimal path is illustrated in Figure 4. Illustrated routes in Figure 3 are more than just simple plots, they powerfully convey the inherent difficulty of the TSP and demonstrate how GTA manages complexity, transforming a difficult combinatorial problem into a solvable, exact, and aesthetically pleasing solution.

The route visualization feature of our GTA method elucidates node clustering behavior and spatial distribution for all instances and number of nodes. Optimization efficiency is indicated in Figure 4 by the network's clustering graphs for N = 1000 and N = 2000 nodes, as the clustering gets denser, visualizing TSP's optimal partitioned-space is an asset, supported as well by the consistent visualization of spatial spacing between the starting and halfway nodes, and the effort expended to provide high node-density maps. Logistics and data scientists, and potentially astrophysicists and biologists, are just a few of the users who can benefit practically from TSP route visualizations. These visual aids make it easier for users to comprehend difficult route optimization intuitively. When N = 10, for example, the clear path facilitates shortest-path planning. At N = 100 and higher, GTA's output provides structured insight into the application of clustering and region-based optimization techniques. This is especially useful in real-world applications such as network layouts, delivery route planning, and various other tracking applications where visual detection of spatial grouping reduces travel distance and increases efficiency. This allows users to experiment with different node configurations and see how routes change, providing them with hands-on visual experience of problem scales, solution complexities, and spatial data behavior, bridging the gap between complex algorithms and real-world application.

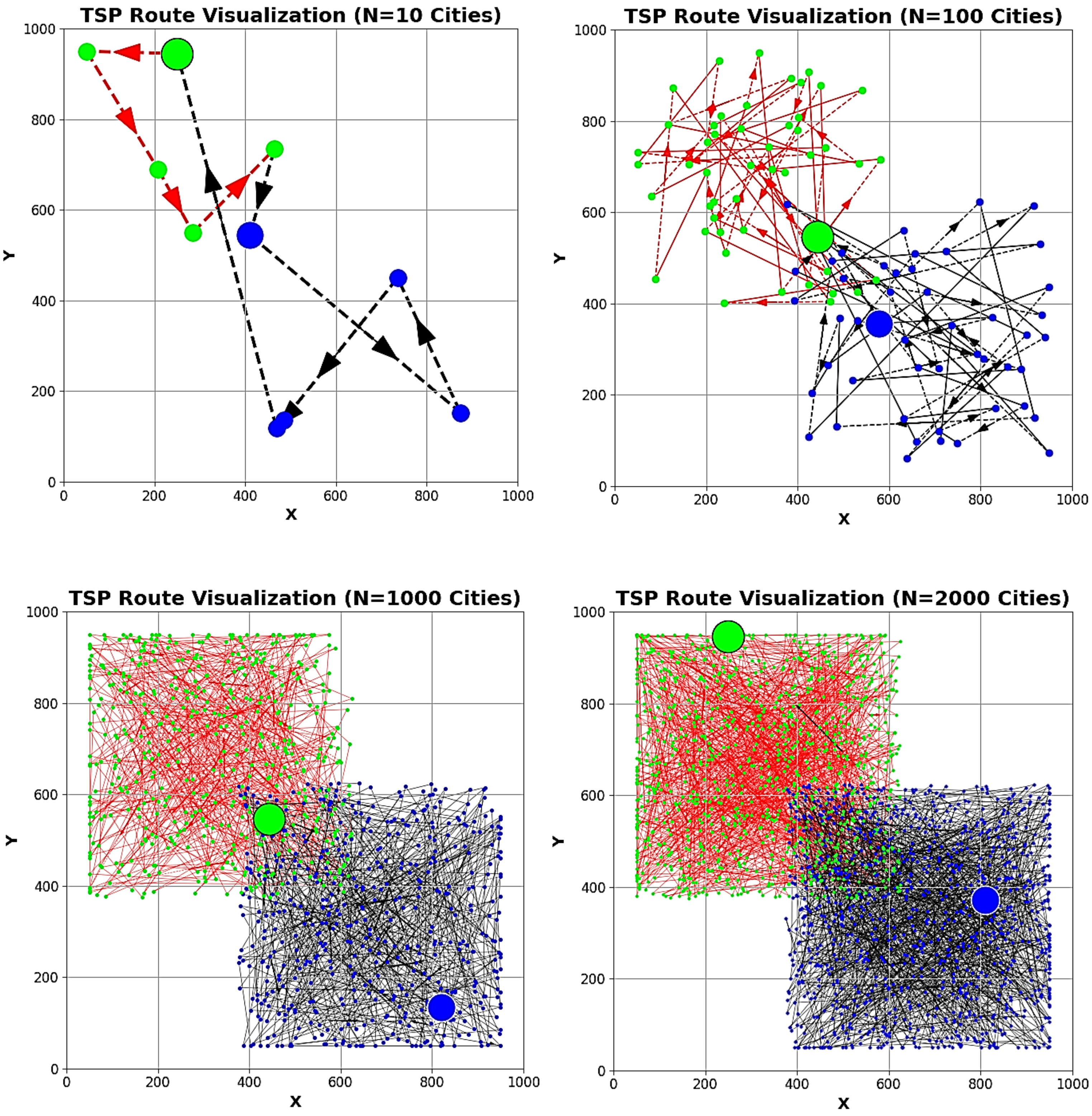

**Figure 4: Optimal Route Maps for Different Node Sizes (The Large Green and Blue Dots, Represent Starting And Mid-Way Nodes)**

**SEEDING & SCALING**

Our GTA algorithm is a two-stage optimization process that uses Gurobi, as a Mixed Integer Programming solver, and Tabu Search to generate a high-quality initial solution, significantly trimming infeasible and sub-optimal search spaces, and critically reducing Gurobi's solution search, resulting in exceptionally fast runtimes. This initial Tabu search allows Gurobi to avoid local optima, making its solution exploration and computing memory less exhaustive. GTA's generalizability is based on its ability to handle both sparse and dense TSP versions (A reminder that all our results are for asymmetric problems ASTP). GTA has proven robustness, and by changing the initial seed value that generates randomized cost/distance matrices, our algorithm's consistency in runtime performance is further demonstrated. Changes in the seed value (i.e., in the cost matrix coefficients) can significantly alter a solver's search space and global optima, and adding new local minima. By showing that GTA maintains the same runtime characteristics when the seed value is changed (e.g. from S = 42 to S = 65), results further validate GTA's consistency, efficiency, stability, and ability to outperform despite input perturbations, i.e. changing the cost matrix. While we show results for two seed values, we have validated GTA's consistency over many other seed values (over 5 values and over the full range of node sizes), our experimental results provide empirical evidence supporting GTA's consistency and resilience, no matter how varied coefficients of the cost matrix are. A comparison of GTA's performance under two different random seeds (S = 42 and S = 65) across a full range of node sizes (N = 10 to N = 4500) is displayed in Figure 5. When the seed is changed, altering the cost matrix coefficients, there is no discernible runtime difference. This consistent runtime demonstrates GTA's predictable computational behavior as the number of nodes increases, for which runtimes converge even more closely (See Figure 5). This randomness invariance, and unwavering runtime performance of our GTA algorithm for different seed values (S = 42 and S = 65) highlights its robustness, scalability, and universal applicability. GTA's runtime seed-invariance strengthens its reliability and predictability, especially for large node counts, with runtimes nearly insensitive to changes and randomization of the initial cost matrix coefficients. As shown in Figure 5, GTA produces predictable, seed-invariant runtime stability, logarithmic consistency, and performance convergence. Figure 5 supports the claim that GTA is a consistent algorithm rather than a statistical anomaly; our experimental data shows that GTA's performance is steady and predictable. This performance consistency and uncontested speed, which is rare for exact deterministic methods, demonstrates a high level of engineering skills in GTA's framework design. GTA is a solution that can be used by engineers and scientists, as well as students and researchers, and can be used without advanced programming skills. GTA can now be positioned in the TSP literature as a unique and benchmark-worthy algorithm.

Variations in the coefficients of the cost matrix could lead to unexpectedly longer runtimes and poor performance. GTA's performance, however, is remarkably unaffected by this variance. Rather than being designed to a specific set of cost coefficients, GTA performs equally well regardless. Thus, our hybrid approach is ideal for real-world applications where cost matrices vary from run to run due to data updates. For smaller node sizes, slight, insignificant fluctuations are observed, but all seed curves converge for larger instances. This predictability and reliability can be expected regardless of the input's randomness, providing users with confidence when using our GTA. Data for three additional seed values (133, 29, and 7) with testing extended up to N = 1000 are presented in Figure 5, to consolidate confidence in the robustness of GTA. All five curves in Figure 5 behave similarly, maintaining nearly the same runtime scaling, which indicates algorithmic consistency under a variety of input conditions. As a reminder, all cases discussed in this work are based on ATSP cost coefficients, making GTA even more reliable, and not selectively or coincidentally good only for specific cases. This gives GTA a further advantage over other exact and heuristic methods that require precise calibration for every problem instance, due to their inability to handle variances in cost coefficients.

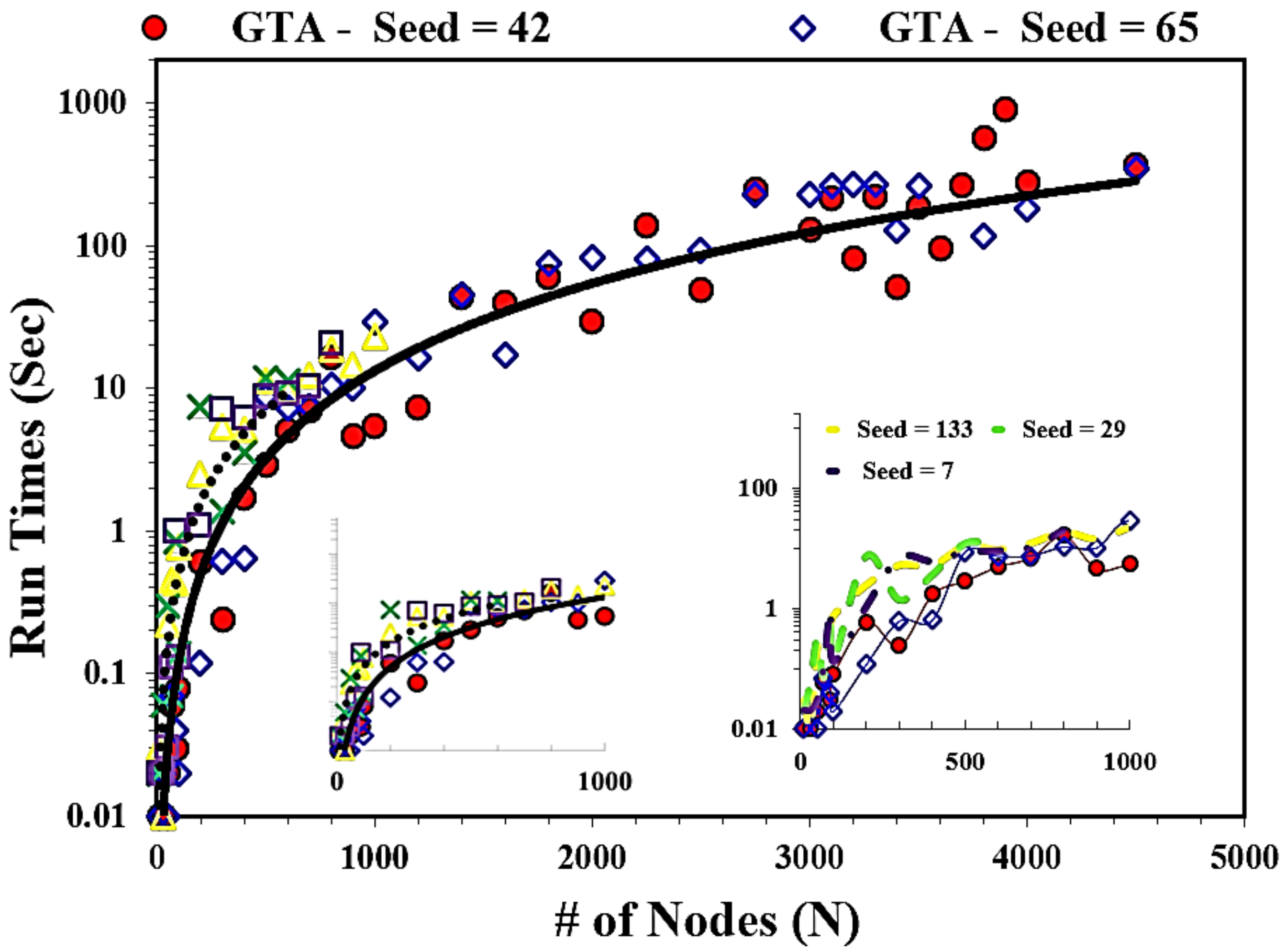

**Figure 5: GTA's Runtime Invariance Across Randomized TSP Instances (Seeds: 42, 65, 7, 29, 133)**

In this next section, we examine GTA's practical and reliable performance in solving combinatorial ATSPs to optimality, by scaling the coefficients of the cost matrix by a factor of 10. We verify GTA's unwavering performance even when the possible values of the cost matrix's random coefficients are critically increased, we widen the numerical range of edge weights by an order of magnitude, effectively scaling them by a factor of 10. It is important to emphasize that this scaling analysis is not mathematically critical. Just as we increase the range of cost coefficients by a factor of 10 (from a range of [1, 10] to [10, 100]), we could equivalently have divided the larger range by 10 and normalized it. Coefficients taken from real problems, regardless of their scale or units, could be readily normalized, by dividing them by a constant factor.

Theoretically, such modifications would change the structure of ATSP problems and affect the search space. Therefore, the objective is to confirm the numerical robustness of our GTA algorithm, and to demonstrate that GTA solves to 0% gaps and in unprecedented runtimes consistently, regardless of the magnitude of the input costs, even when those cost coefficients span larger numerical ranges. This comes as an essential validation, and a crucial reassurance to users, that GTA does not require additional preprocessing, and is immune to larger coefficient ranges and to numerical unpredictability. Two datasets are plotted against the number of nodes ($N$) on a log-log axis in Figure 6, red circles show GTA performance over ATSP matrices generated with cost coefficients in the [1,10] range, and blue crosses show runtimes over cost matrices in the [10,100] range. Both show nearly linear fits on the log-log scale, indicating that the computational complexity of the algorithm remains constant regardless of the coefficients' range. Figure 6 supports our claim that GTA is runtime invariant to larger cost matrix coefficients and a larger coefficient range. The two inset plots are used to reiterate GTA's robustness, a Log-Lin Plot with a sub-exponential growth profile, with the red and blue curves almost indiscernible. The Lin-Lin plot not only illustrates the increasing computing effort with the increase in problem size, but also that curves closely overlap, demonstrating consistent scaling of GTA with respect to cost ranges. Figure 6 is a benchmark and visual evidence of GTA's runtime stability against alterations in the range and values of the cost matrix coefficients. Our GTA solver is immune to statistical random cost changes, all random matrices yield similar results even though their numerical scales are fundamentally different.

Even in cases where cost coefficients differ significantly, GTA retains a data-neutral performance, which may be of interest to users with interest in integrating the GTA solver into larger software systems or supercomputing/parallel clusters. Figure 6 shows that there is no need to normalize cost matrices or change GTA's parameters. All components of our GTA's hybrid approach contribute to the algorithm's independence from cost coefficient scales and range. Given its robustness and stability across different datasets, GTA is a complete and deployable algorithm in various interdisciplinary applications. Figure 6 provides perhaps the best evidence that GTA provides neutral and independent solutions to ATSP problems. Unlike naive solvers or metaheuristics, which suffer from significant performance degradations as the input changes, GTA offers a stable runtime that is almost insensitive to input cost coefficients. GTA is an algorithm that consistently solves ATSP, its consistency is equally important as its unparalleled speed, as it enables obtaining trustworthy solutions that can be integrated across a range of applications.

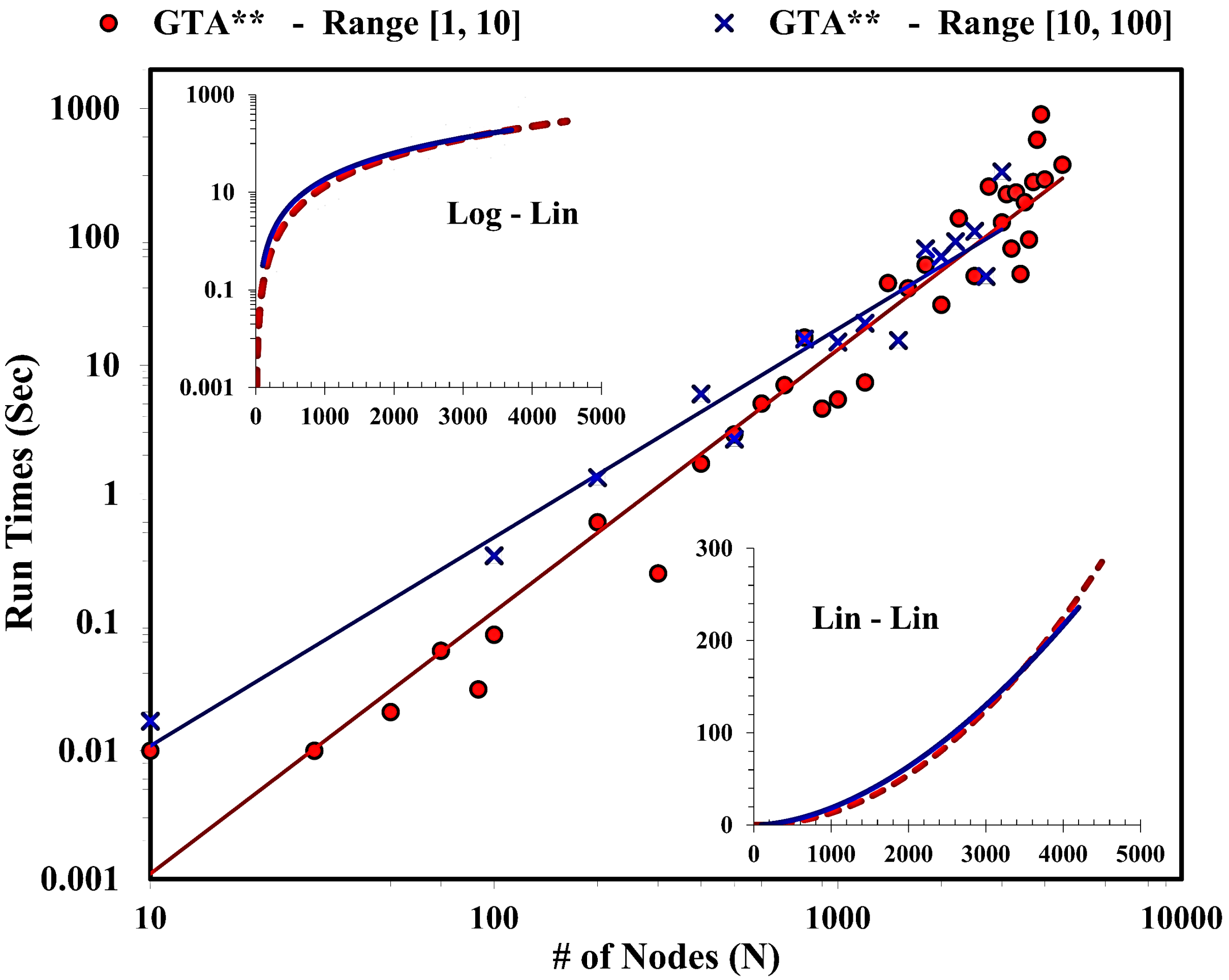

**Figure 6: Runtime Stability of GTA Across Cost Ranges [1, 10] and [10, 100] in ATSP**

## DISCUSSION

In this work, we have introduced GTA: a scalable hybrid method (Gurobi w/ Tabu warm start) for solving large-scale asymmetric Traveling Salesman Problems (ATSPs). We show that GTA consistently achieves optimal solutions and 0% optimality gaps within unprecedented runtimes, even as problem size increases. In fact, for instances larger than N = 4000 nodes, GTA converges to bounded runtimes well below 900 seconds (~15min), and its performance becomes constrained by availability of Random-Access Memory (RAM) and system memory, not computational complexity. The experiments were conducted on a standard 16 GB RAM setup without GPU or parallelization, on an Intel machine with 4 Cores and 8 Processors, with results validated on a 12$^{th}$ generation i5 machine with 8 cores and 16 processors, both with 16GB RAM. We have shown that GTA's runtime becomes horizontally asymptotic for larger instances, due to GTA's hybrid architecture, which reduces Gurobi's search space with a superior Tabu Search warm start, indicating that GTA overcomes computational bottlenecks, contrary to traditional methods. We have also verified that all our results can be replicated down to 1 second accuracy using the same seed values, cost matrix generating methods, and solver parameters. Runtime, solution quality, and solver statistics are explicitly logged. We are dedicated to transparency and are willing to provide full solver logs, command-line prompts, and runtime metadata upon request, including Gurobi's internal metrics, demonstrating our rigorous and unparalleled scientific rigor.

We are also willing to provide our tested GTA variant that includes time window constraints, which are crucial in scheduling telescopes and logistics. GTA is readily adjustable for other variants such as multi-agent TSPs and gene overlap sequencing, but we have not designed and tested such variants. Strategic heuristic initialization, constraint modeling, and solver configurations are kept apart to allow for ease-of-integration of new constraints without interfering with the main algorithm design. GTA’s parameters are simple and require minor adjustments. We only needed to determine the optimal set-up, and then the same nearly identical Gurobi parameters were used for all instances and node sizes, all seed values and cost matrices. Gurobi parameters were kept constant throughout all GTA runs (excluding a few exceptions). Users can experiment with different seed values and problem sizes thanks to its GUI implementation, which is user-friendly and intuitive. GTA is appropriate for most real-world applications due to its seed-invariant behavior, cost-range independence, and consistently short and predictable runtimes.

## CONCLUSION

The Gurobi Tabu (GTA) algorithm is a unique hybrid method considered a giant leap forward in solving Asymmetric Traveling Salesman Problems (ATSP). GTA has overcome various theoretical and computational barriers faced by common methods. GTA provides zero gap and scale-independent solutions in unparalleled runtimes, surpassing traditional methods, especially for big instances of up to 5,000 nodes. Our most innovative and impactful contribution is showing that our GTA algorithm is able to narrow down large and complex solution spaces, and solve large instances of ATSPs to optimality, with runtime approaching a near-constant upper bound, implying that the primary bottleneck for GTA is no longer computational time, but rather hardware limitations: system RAM. The GTA framework is completely open and adaptable and can be quickly modified to adapt to other TSP variants and various combinatorial optimization problems. We are committed to transparency and are eager to share detailed codes and further experimental data, providing a foundation for future work.

## ACKNOWLEDGMENTS

We would like to express our sincere gratitude to everyone involved in this research effort. We are especially thankful for the support from the Rampage Hardware Network, for allowing us access to computing resources to confirm our countless experiments and GTA test runs discussed in this paper. This support was crucial to venturing beyond the usual problem scale and demonstrating the efficiency of our algorithm. Our deepest appreciation goes to Dr. Roger Ashkar, whose support for this project was unwavering, providing a solid foundation for it. Finally, we acknowledge the valuable contribution of Alumni/Graduates Abbas Toufaili and Mostafa Tormos, whose input for this work was essential to its development.

---